# Electronic Properties of BaPtP with a Noncentrosymmetric Cubic Crystal Structure

Yoshihiko Okamoto[1,*], Ryosuke Mizutani[1], Youichi Yamakawa[2], Hiroshi Takatsu[3], Hiroshi Kageyama[3], and Koshi Takenaka[1]

[1]*Department of Applied Physics, Nagoya University, Nagoya 464-8603, Japan*
[2]*Department of Physics, Nagoya University, Nagoya 464-8602, Japan*
[3]*Graduate School of Engineering, Kyoto University, Kyoto 615-8510, Japan*

We report the synthesis, electronic properties, and electronic structure of LaIrSi-type BaPtP with a noncentrosymmetric cubic crystal structure. Electrical resistivity and heat capacity data taken by using polycrystalline samples indicated that BaPtP is a metal, which was further supported by first principles calculations. A polycrystalline sample of BaPtP showed a zero resistivity below 0.2 K due to the superconducting transition. The first principles calculation results indicated that the spin splitting at around the Fermi energy is large in BaPtP. These results suggest that BaPtP is likely to exhibit interesting physical properties caused by a strong spin-orbit coupling of 5*d* electrons in the Pt atoms.

## 1. Introduction

Platinum-based superconductors have been intensively studied in recent years as a platform where unconventional superconductors can be realized due to the strong spin-orbit coupling and the orbital degree of freedom of 5*d* electrons, with typical examples including $Li_2Pt_3B$ and YPtBi [1,2]. The spin-triplet pairing was reported to be dominant in $Li_2Pt_3B$ [3,4], while the possibility of topological superconductivity caused by the $J = 3/2$ state was discussed in YPtBi [5-9]. Common to both compounds, the crystal structure is cubic and does not have space inversion symmetry, which are both expected to play an important role in the realization of unconventional superconductivity. Very recently, ullmannite-type PtSbS was reported to be a superconductor with a transition temperature of $T_c = 0.15$ K [10]. The ullmannite-type structure has the cubic $P2_13$ space group without space inversion symmetry. First principles calculations indicated that the Fermi surfaces of PtSbS include strongly nested hole pockets, which might be related to the emergence of superconductivity, although the spin splitting at around the Fermi energy is small.

In this study, we report the physical properties of a noncentrosymmetric cubic Pt compound, BaPtP, which has the $P2_13$ space group, the same as for PtSbS. BaPtP was first synthesized by G. Wenski and A. Mewis and reported to crystallize in the LaIrSi-type structure [11], as shown in Fig. 1(a) [12]. In this crystal structure, each P and Pt atom forms a trillium lattice, where each corner of $P_3$ and $Pt_3$ regular triangles is shared by three $P_3$ and $Pt_3$ triangles, respectively [13]. This crystal structure can also be interpreted as an anti-ullmannite type. There is no previous report on the physical properties of BaPtP. In *A*Pt*X* compounds, where *A* and *X* are alkali earths and pnictogens, respectively, only BaPtP and a polymorph of BaPtAs crystallize in the LaIrSi-type. The LaIrSi-type BaPtAs was reported not to show a superconducting transition down to 0.1 K [11, 14]. This is in contrast to the *A*Pt*X* family with a honeycomb structure of Pt and $X$ atoms, in which many compounds, such as SrPtAs and BaPtSb, show superconductivity with $T_c$ = 2–3 K [14-16]. In the LaIrSi-type compounds, EuPtSi was intensively studied in terms of its magnetic properties. This compound is an antiferromagnet with a Néel temperature of 4 K [17, 18], and was found to show various magnetic phases in magnetic fields [19, 20].

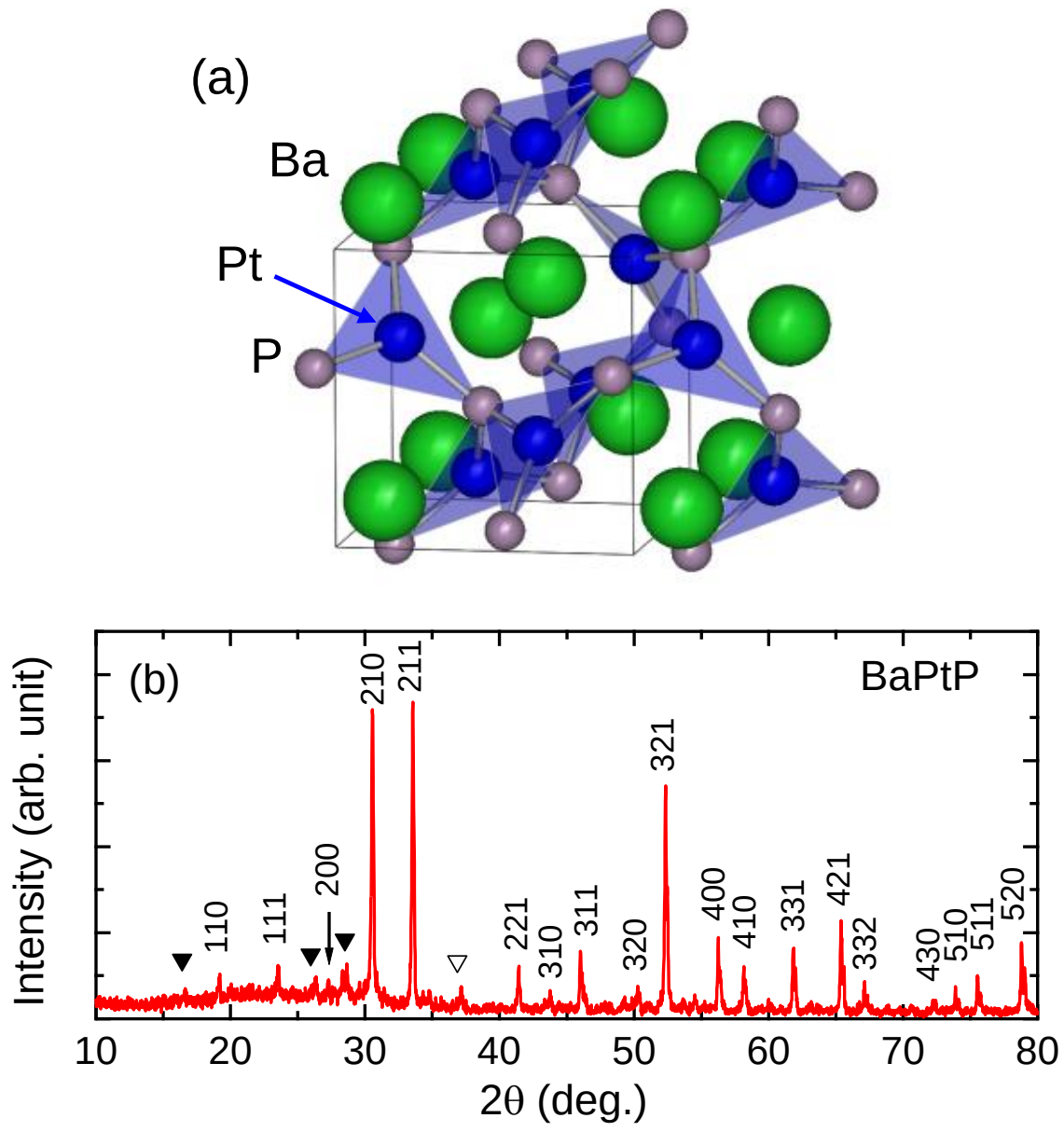

Fig. 1. (a) Crystal structure of BaPtP. The solid line indicates the unit cell. (b) A powder X-ray diffraction pattern of a BaPtP polycrystalline sample taken at room temperature, indexed by a cubic cell of $a = 6.533$ Å. The peaks indicated by filled and open triangles are those from $BaPt_2P_3$ and unknown impurities, respectively.

## 2. Experimental Procedure

Polycrystalline samples of BaPtP were prepared by a solid-state reaction method. First, Ba powder was obtained by filing Ba shots (99%, Rare Metallic) in an inert gas atmosphere. A mixture of the Ba powder, Pt powder (99.9%, Rare Metallic), and black phosphorous (99.9999%, Kojundo Chemical Lab.) of a 1.2 : 1 : 1 molar ratio was pressed and then sealed in an evacuated quartz tube. The tube was first kept at 673 K for 24 h to avoid rapid vaporization of phosphorous. Next, the tube was kept at 1273 K for 72 h, and then furnace-cooled to room temperature. The obtained sample was grinded, pressed, and then sealed in an evacuated quartz tube again. The tube was kept at 1273 K for a few days. Sample characterization was performed by a powder X-ray diffraction (XRD) analysis with Cu K$\alpha$ radiation at room temperature using a RINT-2100 diffractometer (Rigaku). As shown in Fig. 1(b), all diffraction peaks, except some small peaks due to minute amounts of $BaPt_2P_3$ and unknown impurities, were indexed on the basis of a primitive cubic unit cell of $a$ = 6.533 Å, indicating that the cubic BaPtP phase was obtained as a main phase. The electrical resistivity, heat capacity, and magnetization of BaPtP polycrystalline samples down to 2 K were measured using Physical Property Measurement System and Magnetic Property Measurement System (both Quantum Design). Electrical resistivity measurements down to 0.1 K were performed using an adiabatic demagnetization refrigerator. First principles calculations were performed using the WIEN2k code [21]. The experimentally obtained structural parameters were used for the calculations [11].

## 3. Results and Discussion

### *3.1 Electronic properties*

Figure 2 shows the temperature dependences of electrical resistivity, ρ, of a BaPtP polycrystalline sample. The data for PtSbS are also shown as a reference [10]. With decreasing temperature, ρ of BaPtP decreases, indicative of the metallic behavior. The residual resistivity is approximately 10 μΩ cm, which is much smaller than that of PtSbS. At lower temperatures, ρ drastically decreases below 0.22 K and becomes zero at 0.2 K, as shown in Fig. 2(b). The zero resistivity temperature decreases by applying magnetic fields, suggesting that this resistivity drop is caused by a superconducting transition. It should be confirmed whether the transition is a bulk transition or not by heat capacity or magnetization measurements, which would establish BaPtP as the second superconducting Pt compound with the $P2_13$ space group, in addition to PtSbS.

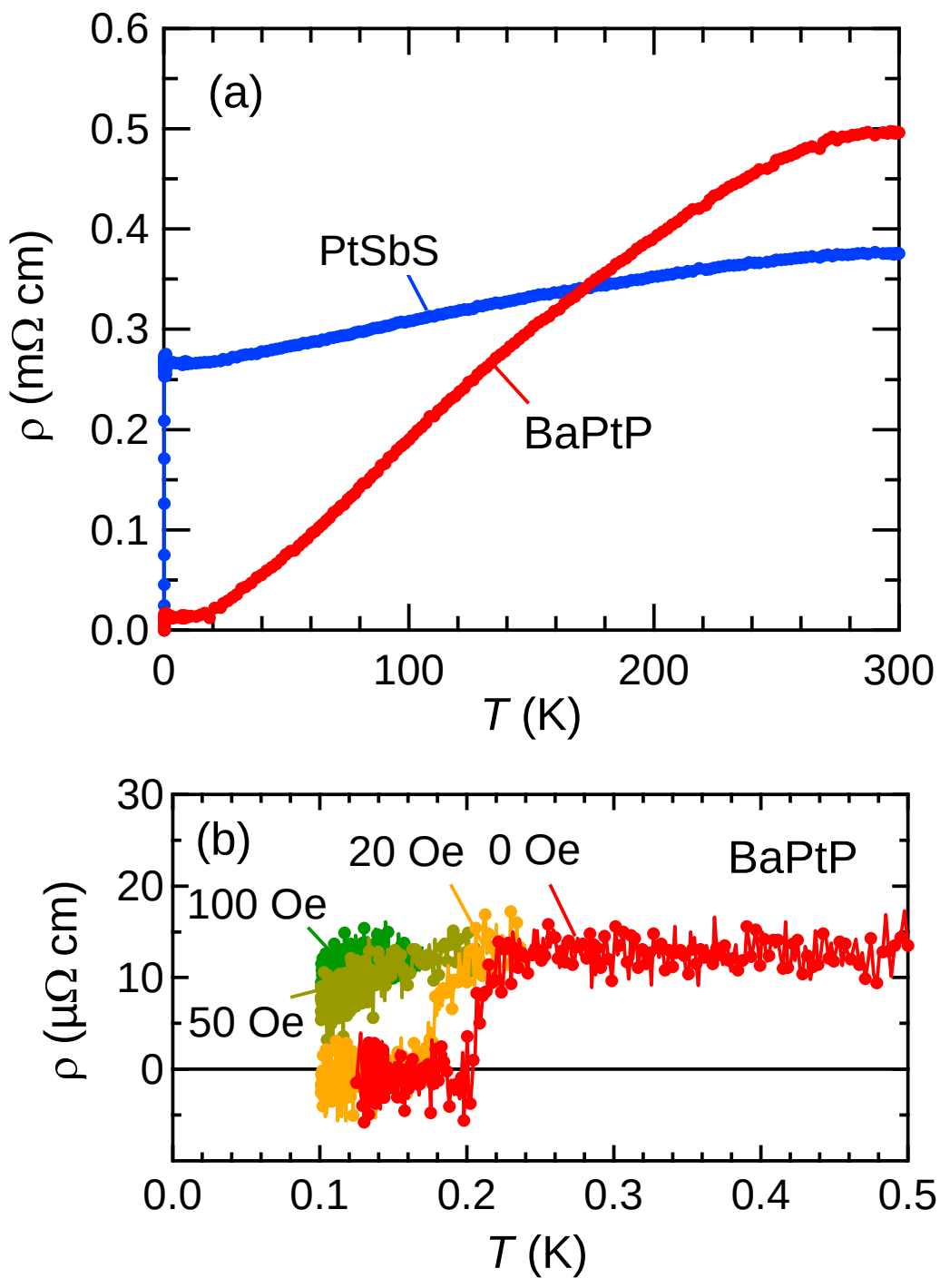

Fig. 2. (a) Temperature dependence of electrical resistivity of a BaPtP polycrystalline sample. The data of PtSbS are also shown as a reference [10]. (b) Temperature dependence of electrical resistivity of a BaPtP polycrystalline sample measured at various magnetic fields of 0–100 Oe.

A critical magnetic field of this superconducting phase is small. As shown in Fig. 2(b), the midpoint and zero resistivity $T_c$ are suppressed by applying a magnetic field of 50 Oe, suggesting that the critical magnetic field is less than 100 Oe. This critical magnetic field is much smaller than 620 Oe in PtSbS ($T_c$ = 0.15 K) and comparable to those of type-I superconductors. It would be interesting if BaPtP is a type-I superconductor, because almost all superconductors except for the elemental superconductors are type-II [22].

Figure 3 shows heat capacity and magnetic susceptibility of BaPtP. The heat capacity data shown in Fig. 3(a) yield $\beta$ = 0.499(4) mJ $K^{-4}$ $mol^{-1}$ and an electronic specific heat of $\gamma$ = 2.70(3) mJ $K^{-2}$ $mol^{-1}$ by a linear fit to $C/T = \beta T^2 + \gamma$ between 2.0 and 3.2 K. This $\beta$ value yields Debye temperature of $\theta_D$ = 227 K. The finite $\gamma$ value is consistent with the metallic behavior of the electrical resistivity shown in Fig. 2(a). The magnetic susceptibility of BaPtP shown in Fig. 3(b) gradually decreases with increasing temperature and is almost zero above 30 K, implying that the Pauli paramagnetic and Larmor diamagnetic contributions almost cancel each other out.

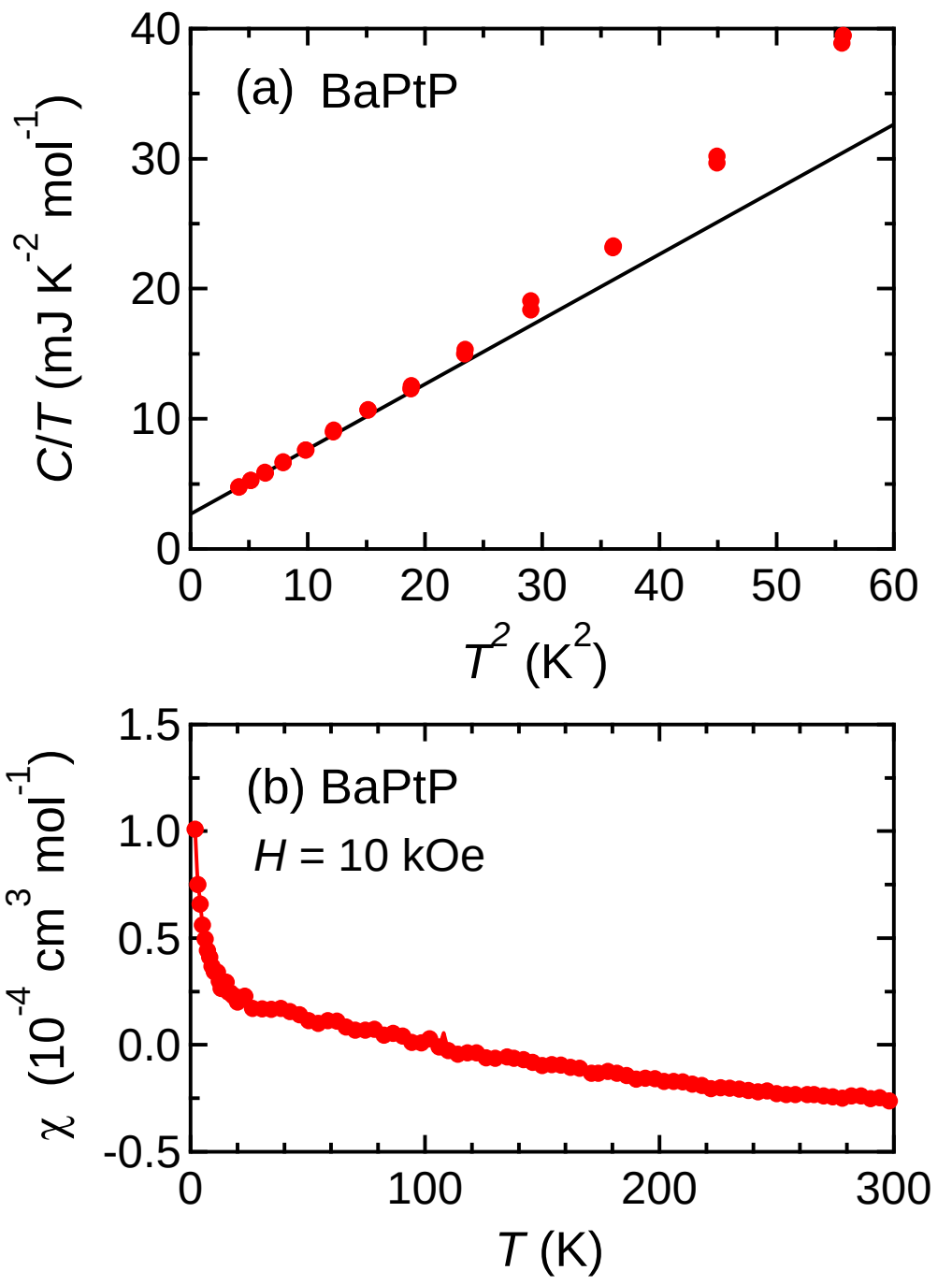

Fig. 3. (a) Heat capacity divided by temperature of a BaPtP polycrystalline sample as a function of $T^2$. The solid line shows the linear fit between 2.0 and 3.2 K. (b) Temperature dependence of magnetic susceptibility of a BaPtP polycrystalline sample measured at a magnetic field of 10 kOe.

### *3.2 Electronic structure*

We now discuss the electronic structure of BaPtP. The electronic band structure and electronic density of states (DOS) of BaPtP calculated with spin–orbit coupling are shown in Fig. 4. The electronic bands cross the Fermi energy $E_F$, indicative of the metallic band structure, which is consistent with the observed metallic ρ. As seen in the electronic DOS shown in the right panel of Fig. 4, the contribution of Pt at $E_F$ is considerably larger than those of Ba and P. The electronic specific heat of BaPtP is calculated to be $\gamma_{band}$ = 2.0 mJ K$^{-2}$ mol$^{-1}$ from the DOS at $E_F$ (~3.4 eV$^{-1}$), which is ~40% smaller than the γ determined by the heat capacity data, suggesting that γ is enhanced by electron-electron and/or electron-phonon interactions.

A characteristic point of the electronic structure of BaPtP is considerably large spin splitting in the energy bands at around $E_F$, which is caused by an antisymmetric spin–orbit coupling. As seen in Fig. 4, the splitting on X–X' and Γ–X lines at around $E_F$ is 0.1–0.2 eV, much larger than those in PtSbS [10]. This result means that the strong spin–orbit coupling of Pt 5*d* electrons plays an important role in the spin splitting in BaPtP, in contrast to the case of PtSbS. In BaPtP, the spin-split bands cross $E_F$ on the X–X' line, resulting in small Fermi pockets, while the bands are located just below $E_F$ on the Γ–X line, suggesting that a spin-polarized Fermi pocket made by one of the spin-split bands can be formed by chemical substitution or applying a pressure.

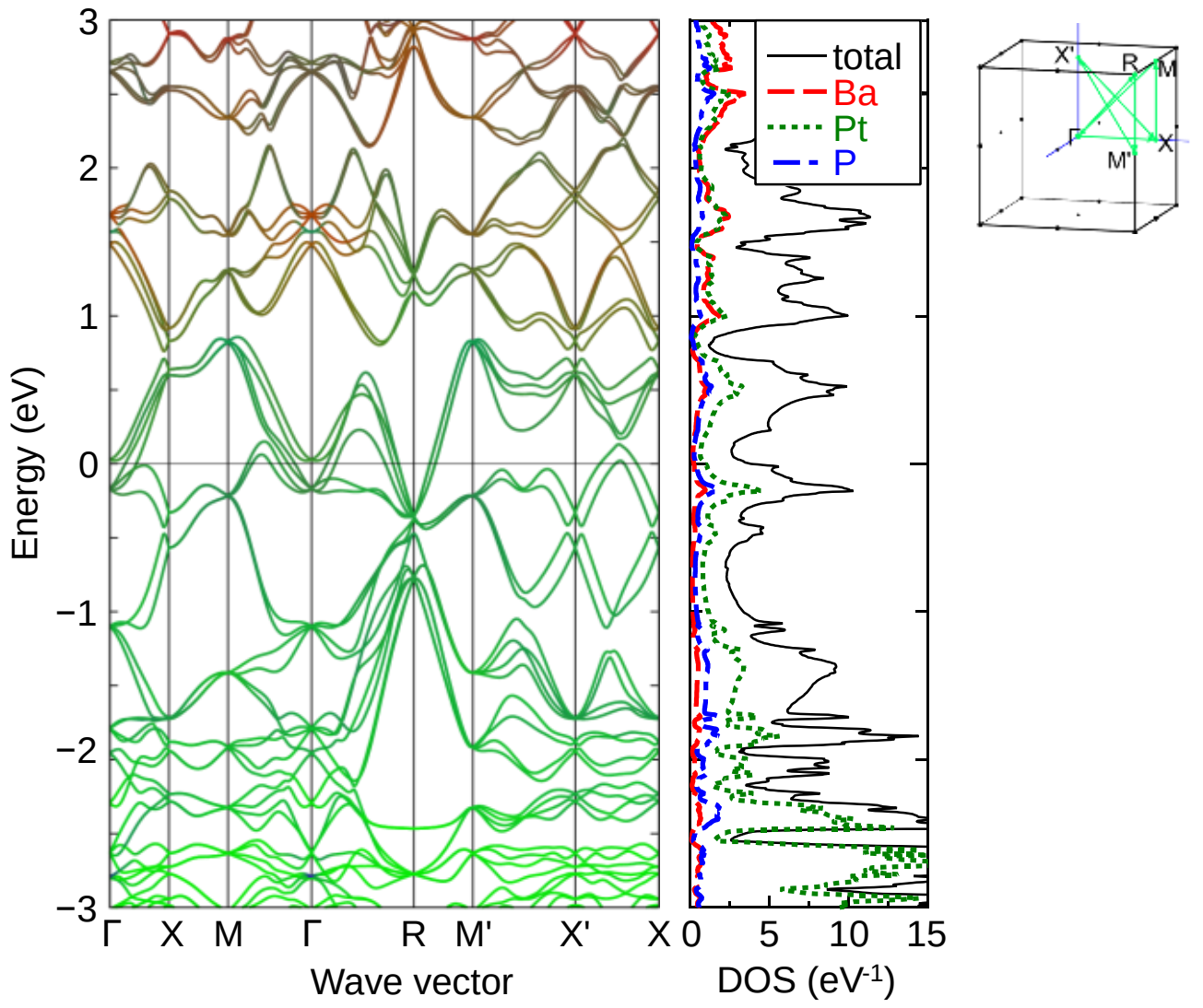

Fig. 4. Electronic structure of BaPtP calculated with spin–orbit coupling. Electronic band structure, total and partial DOS, and the first Brillouin zone are shown. The Fermi level is set to 0 eV.

## 4. Conclusion

LaIrSi-type BaPtP possessing a cubic crystal structure without space inversion symmetry was found to show metallic electronic properties. A polycrystalline sample of BaPtP showed zero resistivity below 0.2 K due to a superconducting transition with a very small critical magnetic field up to 100 Oe. The bulk superconductivity in BaPtP should be confirmed by future heat capacity or magnetization measurements. First principles calculations results indicated that the spin splitting by the antisymmetric spin-orbit coupling is considerably large in BaPtP. These results suggest that BaPtP is promising to show interesting physical properties caused by the strong spin-orbit coupling in 5*d* electrons of Pt atoms.

## Acknowledgments

This work was partly carried out under the Visiting Researcher Program of the Institute for Solid State Physics, the University of Tokyo and supported by JSPS KAKENHI (Grant Number: 18H04314, 19K21846).

*E-mail: yokamoto@nuap.nagoya-u.ac.jp